\documentstyle[12pt]{iopart}
\begin{document}
\title [Templates for Smirches] 
{Templates for stellar mass black holes falling into supermassive black holes}
\author{B.S. Sathyaprakash$^1$ and B.F. Schutz$^{1,2}$}
\address{$^1$Department of Physics and Astronomy, Cardiff University, Cardiff, UK
and $^2$Albert-Einstein Institute, Golm, Germany}
\begin{abstract}
The spin modulated gravitational wave signals, which we shall
call {\it smirches}, emitted by
stellar mass black holes tumbling and inspiralling into massive
black holes have extremely complicated shapes. Tracking these
signals with the aid of pattern matching techniques, such
as Wiener filtering, is likely to be computationally an impossible
exercise. In this article we propose using a mixture of optimal
and non-optimal methods to create a search hierarchy to ease
the computational burden. Furthermore, by employing the
method of {\it principal components} (also known as {\it singular
value decomposition}) we explicitly demonstrate that the effective
dimensionality of the search parameter space of smirches is likely to be 
just three or four, much smaller than what has hitherto been thought
to be about nine or ten. This result, based on a limited study
of the parameter space, should be confirmed by a more exhaustive
study over the parameter space as well as Monte-Carlo simulations
to test the predictions made in this paper.
\end{abstract}
\pacs{04.80.Nn, 95.55.Ym, 95.75.Pq, 97.80.Af}
\maketitle
\submitto{\CQG}

\section{Introduction}
A stellar mass compact object, either a neutron star or a black hole, in tight orbit
around a massive Kerr black hole will tumble around the massive companion
due to strong spin-orbit coupling. For small mass ratios the stellar mass companion
{\it slowly} inspirals into the massive hole, the duration of inspiral increasing 
inversely as the mass ratio. Gravitational waves emitted in the process of 
adiabatic inspiral encode the structure of the  spacetime geometry around the hole
thereby allowing us to test the uniqueness theorems of black holes spacetimes \cite{RyanAndFT00}.
However, decoding the information carried by the waves would require an accurate
knowledge of how the emitted waves depend on the parameters of the system and 
fitting the data with the expected signal. There are problems with
both of these aspects: On the one hand we do not currently have an accurate knowledge
of the signal although over the next decade the situation is likely
to have improved. On the other hand, the problem of fitting is unlikely to go away
as the number template waveforms required to decode the useful information might 
be formidably large. This is because the dimensionality of the parameter space is
quite large. Indeed, in the generic case of a binary consisting of two spinning
black holes there are potentially seventeen parameters to fit: The direction and
distance to the source, two masses, three vectors corresponding to the initial values 
of the orbital angular momentum and two spin angular momenta, the eccentricity 
of the orbit, the instant when the orbital frequency reaches a fiducial value 
and the phase of the signal at that instant. 

Even though the time-structure of the signal depends on seventeen parameters,
varying different parameters may not always produce distinct waveforms. In other words,
the effective dimensionality of the parameter space for the purpose of fitting
might be a lot smaller.  For instance, we know in the case of non-spinning 
binaries that out of potentially eight parameters -- namely, the direction to the
source and its distance from the Earth, two masses, an initial magnitude of 
the angular momentum, a fiducial time and phase when the signal reaches
a certain frequency -- there is effectively only
one parameter with which to fit the data. This is the so-called {\it chirp mass,}
which is a certain combination of the two masses of the component bodies. The
detector response has no time-varying dependence on the source direction
since the signal lasts for a very short duration (minutes, at most) and 
therefore doesn't suffer  any modulation due to the motion of the antenna that 
could have been useful in deciphering the source direction; the initial magnitude 
of the angular momentum only sets an initial frequency of observation that 
is chosen by hand;  the fiducial time and phase can be easily determined by 
the use of the fast Fourier transform and analytically, respectively; and 
the phase of the signal, which is what is important in accomplishing a good fit, 
depends, at the dominant post-Newtonian order, only on the chirp mass. 
Although the degeneracy in component masses is resolved
when higher order post-Newtonian terms are included in the phasing, it turns 
out the number of `independent' shapes doesn't increase significantly while 
including a second mass parameter. 

An analysis similar to the non-spinning case is required in the case of spinning
black hole binaries. In this paper we discuss the method of {\it principal components} 
\cite{BSS98} that can be used to determine the effective dimensionality of the search
space. We will then apply it to the case of a test mass falling into a massive
black hole. We sample the parameter space at a few crucial points to get an
idea of the number of
principal components on which the signal shape depends, thereby obtaining
an estimate of how large the parameter space could be. We begin with a 
rough counting of the number of template waveforms that may be required in
a fitting problem. This is really an upper limit on the number of waveforms,
the actual number might be a lot less.

Since we expect the number of templates to be too
large it will not be possible to employ matched filtering on an entire
year's worth of data. Rather, the idea is to employ a technique, such as 
time-frequency analysis, that makes little use of the signal shape in 
capturing a signal, albeit with a lower detection probability. 
Such an analysis might help to first get a rough idea 
about the signal parameters. One could then
employ matched filtering on, say a week's worth of data and
refine the estimation of parameters. Based on these refined parameters
one finally employs the search
on a year's worth of data but in a small region of the parameter space.
In other words, we are advocating a multi-step hierarchical search involving
different analyses techniques. In such a scheme it would only be necessary
to estimate the number of templates required in the most compute intensive
stage of the search hierarchy, namely in the first stage of matched filtering.
It is best to begin the search from a point where the parameters can be nailed
down most easily, namely by employing matched filtering on a week's worth of
data. (At this stage a signal to noise ratio of about 5-10 can be expected
for mergers taking place at a distance of about a few Gpc).  This assumption
somewhat simplifies our calculation as we can neglect the motion of LISA
relative to the source during this period and assume that the detector
is fixed relative to the source. Strictly speaking, this is not a valid
assumption but it is unlikely to change the qualitative conclusions reported
in this work.

\section{Counting the number of templates -- A rough estimate}
To count the number of templates we must first understand how the signal
evolves as a function of time. To this end we shall use the best post-Newtonian 
order currently known but ignore eccentricity.  It turns out that at this order 
the signal already has many modulations including spin-orbit and spin-spin couplings.
Therefore, the waveform should be quite suitable for counting template numbers
although it may not be good enough for the purpose of fitting. 

\subsection{Evolution of the orbit}
Following Ref. \cite{acst94} we shall use a 
coordinate system adapted to LISA.  In the post-Newtonian approximation
the evolution of a binary 
system comprising of two bodies of masses $m_1$ and $m_2,$ spins 
${\mathbf S}_1=(s_1,\, \theta_1, \, \varphi_1)$ and 
${\mathbf S}_2=(s_2,\, \theta_2, \, \varphi_2),$ orbital angular momentum
${\mathbf L}=(L,\, \theta_0, \, \varphi_0),$ is governed by a 
set of differential equations given by:
\begin{eqnarray}
\fl \dot{\mathbf {L}} = 
\left [ 
\left ( \frac{4m_1+3m_2}{2m_1m^3} - \frac{3\,  {\mathbf S}_2 \cdot {\mathbf L}}{2\, L^2m^3} \right ) {\mathbf S}_1 + 
\left ( \frac{4m_2+3m_1}{2m_2m^3} - \frac{3}{2}\frac{ {\mathbf S}_1 \cdot {\mathbf L}}{L^2m^3} \right ) {\mathbf S}_2 
\right ] \times {\mathbf L} v^6 \nonumber \\
-\frac{32\eta^2 m}{5L} {\mathbf L} v^7,
\label{eqn:precession1}
\end{eqnarray}
\begin{equation}
\fl
\dot{ \mathbf  S}_1  = 
\left [ \left ( \frac{4m_1+3m_2}{2m_1m^3} - \frac{3\, {\mathbf S}_2 \cdot {\mathbf L} }{2\, L^2m^3} 
\right ) {\mathbf L} \times {\mathbf S}_1 + \frac{ {\mathbf S}_2 \times {\mathbf S}_1 }{2m^3} \right ] v^6,
\label{eqn:precession2} 
\end{equation}
\begin{equation}
\fl
\dot{ \mathbf  S}_2  = 
\left [ \left ( \frac{4m_2+3m_1}{2m_2m^3} - \frac{3\, {\mathbf S}_1 \cdot {\mathbf L} }{2\, L^2m^3} 
\right ) {\mathbf L} \times {\mathbf S}_2 + \frac{ {\mathbf S}_1 \times {\mathbf S}_2 }{2m^3} \right ] v^6.
\label{eqn:precession3} 
\end{equation}
Here $v=(\pi m f)^{1/3},$ where $f$ is the gravitational wave frequency,
is the post-Newtonian velocity parameter, an overdot denotes the time-derivative,
$m = m_1+m_2$ is the total mass, and $\eta = m_1m_2/m^2$ is
the (symmetric) mass ratio. In the evolution of the orbital angular 
momentum we have included the lowest order dissipative term 
[\,the second term containing $v^7$ in Equation (\ref{eqn:precession1})\,] 
while keeping the non-dissipative modulation effects
caused by spin-orbit and spin-spin couplings
[\,the first term within square brackets containing 
$v^6$ in Equation (\ref{eqn:precession1})\,]. 
The spins evolve non-dissipatively but their orientations change due to 
spin-orbit and spin-spin couplings. 
Though the non-dissipative terms are not responsible for
gravitational wave emission, and therefore do not shrink the
orbit, they cause to precess the orbit.

In the absence of spins the antenna observes the same polarization at 
all times; the amplitude and frequency of the signal both
increase monotonically, giving rise to a chirping signal. 
Precession of the orbit and spins cause modulations in the amplitude
and phase of the signal and smear the signal's energy spectrum over a wide band.
The smearing is particularly dramatic
when one of the bodies is much lighter than the other such as the inspiral of a 
stellar mass compact object into a massive black hole. In this case the
orbit would undergo many precessions before the small object plunges into
the hole smearing the signal into hundreds of frequency bins as opposed
to a few frequency bins if the signal were a simple chirp. For this reason
we shall call the spin modulated (sm) chirp, a {\it smirch} (an anagram of 
{\it sm} and {\it chir}).

\subsection{The waveform}
The strain $h(t)$ produced by a smirch in LISA is given by
\begin{eqnarray}
 h(t) & = & -A(t)\cos[\Phi(t)+\varphi(t)],
\label{eqn:waveform}
\end{eqnarray}
where $A(t)$ is the precession-modulated amplitude of the signal,
$\Phi(t)$ is its post-Newtonian carrier phase that
increases monotonically and $\varphi(t)$ is the polarization phase caused 
by the changing polarization of the wave relative to the antenna. (We have
neglected the Thomas precession of the orbit which induces additional, but
small, corrections in the phase.) For a source with position vector 
${\mathbf N} = (D,\, \theta_S,\, \varphi_S)$ the amplitude is given by,
\begin{equation}
\fl
 A(t) = \frac{2\eta m v^2}{D}
\left[ \left( 1 + (\hat {\mathbf L}\cdot \hat {\mathbf N})^2\right )^2 F_{+}^2(\theta_S,\varphi_S,\psi) 
 + 4 \left ( \hat {\mathbf L}\cdot \hat {\mathbf N} \right )^2 F_\times^2(\theta_S,\varphi_S,\psi) \right]^{1/2}. 
\label{eqn:amplitude}
\end{equation}
Here $\hat{\mathbf L} = {\mathbf L}/L,$ $\hat{\mathbf N} = {\mathbf N}/D$ ($D$ is the distance to 
the source) and $\psi(t)$ is the precession-modulated polarization angle given by 
\begin{equation}
\tan \psi(t) = 
\frac{\hat{\mathbf L}(t)\cdot\hat{\mathbf z} - (\hat{\mathbf L}(t)\cdot\hat{\mathbf N})
(\hat{\mathbf z}\cdot\hat{\mathbf N})}
{\hat{\mathbf N}\cdot(\hat{\mathbf L}(t)\times\hat{\mathbf z})}.
\label{eqn:psi(t)}
\end{equation} 
Also, $F_+$ and $F_\times$ are the antenna beam pattern functions are given by
\begin{equation}
F_+(\theta_S,\varphi_S,\psi) = \frac{1}{2}\left(1+\cos^2\theta_S\right)\cos{2\phi_S}\cos{2\psi} 
- \cos\theta_S\sin{2\phi_S}\sin{2\psi}, 
\end{equation}
\begin{equation}
F_{\times}(\theta_S,\phi_S,\psi) = 
\frac{1}{2}\left(1+\cos^2\theta_S\right)\cos{2\phi_S}\sin{2\psi}
+ \cos\theta_S\sin{2\phi_S}\sin{2\psi}.
\end{equation}
Next, the polarization phase $\varphi(t)$ is 
\begin{equation}
\tan \varphi(t) = 
\frac{ 2 \hat {\mathbf L}(t) \cdot \hat {\mathbf N}\, F_\times(\theta_S,\varphi_S,\psi)}
{\left [ 1 + \left ({\mathbf L}(t)\cdot {\mathbf N} \right )^2 \right ] F_{+}(\theta_S,\varphi_S,\psi)}.
\label{eqn:varphi}
\end{equation}
And finally, for the carrier phase we use the post-Newtonian 
expression, but without the spin-orbit and
spin-spin couplings. These spin couplings modify the carrier phase 
by amounts much smaller than the post-Newtonian effects so that neglecting
them makes no appreciable difference to our main conclusions. To second
post-Newtonian order the carrier phase is given by
\begin{equation}
\fl
\Phi(t) = 
\frac{-2}{\eta\Theta^5}\left[1+\left(\frac{3715}{8064}+\frac
{55}{96}\eta\right)\Theta^2-\frac{3\pi}{4}\Theta^3 
+\left(\frac{9275495}{14450688}+\frac{284875}{258048}\eta+\frac{1855}{2048}\eta^2\right)\Theta^4\right].
\label{eqn:Phi}
\end{equation}
where $\Theta=[\eta (t_C-t)/(5m)]^{-1/8},$ $t_C$ being the time at which the two stars
merge together and the gravitational wave frequency formally diverges.

\subsection{A rough count of templates}
From the foregoing equations we can get an idea of the number of parameters that may
be needed in a search. The system we are interested in is that of a stellar mass black
hole falling into a supermassive black hole wherein the mass ratio could be as low
as $10^{-5}$ . Consequently, the spin of the stellar mass hole, being proportional 
to the square of its mass, would be negligible compared to the spin of the massive hole.
Further,  we don't have to specifically search for the distance to the binary nor 
for the fiducial time and phase when the frequency of the emitted gravitational 
wave reaches a certain value as they can all be measured without incurring 
significant computational costs. Moreover, since we are focussing on quasi-circular
orbits (i.e. zero eccentricity) and only those systems that merge within a week
of observation (i.e. after about 1000 orbital time scales) 
the number of parameters in our search is down from 17 to 9. These are
the initial direction of the orbital angular momentum (2 angles), the spin of
the massive black hole (1 magnitude and 2 angles), the masses of the two
bodies (2 numbers) and the direction to the source (2 angles).  

As mentioned in the introduction we don't expect both masses to be important
but only a combination -- the {\it chirp mass} ${\cal M}=\eta^{3/5}m$. 
The azimuth angles are normally not important in a 
search but co-latitudes would lead to modulations that cannot be easily
searched for. Consequently, out of the six angles we expect only three to
be important. (Note that due to different orientations of the binary and
the antenna these angles get mixed up and therefore azimuth
and co-latitudes are not well-defined. However, we expect certain combinations
of these angles to play the roles of azimuth and co-latitude. 
Our claim is that three of those combinations would be important in a 
search and the other three not so important.) In addition, we expect the
spin magnitude of the hole to be a search parameter although this would
be important only when $s_1$ is close to 1. As a result the number of 
search parameters would be five or less. Precisely how many would have
to be determined by a more rigorous and quantitative analysis. Assuming
that we have to search for five parameters in all and that each parameter
requires as many templates as the number of cycles we get a rough count
of $(10^3)^5=10^{15}$ templates. We shall see below that this is a rather
large over estimate. The number of independent parameters is possibly
about three leading to a significant reduction in the number of templates.

\section{Covariance matrix, correlation coefficients and principal components}

Any measurement process is disturbed by a background perturbation or noise
which causes a systematic and/or random errors in the measured quantities.
In general, the errors get smaller at greater signal-to-noise ratios. 
We review the theory behind estimating errors involved in a measurement process.

For a signal buried in a Gaussian background and detected using matched
filtering the covariance matrix $\Sigma_{kl}$ -- a symmetric matrix whose diagonal elements
are the variances in the measurement of the various parameters and the
off-diagonal elements are the covariances between them.
The covariance matrix is itself the inverse of the information matrix, or the
metric,
$g_{mn}.$ Let us suppose the signal $h(t;\lambda_k)$ depends on $p$ 
parameters $\lambda_k,$ $k=1,\ldots,p.$ The information matrix is defined by
\begin{equation}
g_{mn} = \left < \partial_m h\, , \partial_n h \right >,\ \ 
\Sigma_{mn} = \left [ g^{-1} \right]_{mn},
\end{equation}
where for any two functions $a(t)$ and $b(t)$ (and Fourier transforms 
$\tilde{a}(f)$ and $\tilde{b}(f)$) their scalar product 
$\left < a\, , b \right >$ is defined as:
\begin{equation}
\left < a\, , b \right > = 2 \int_{f_{\rm low}}^{f_{\rm high}} \frac{df}{S_h(f)} 
\left [ \tilde{a}(f) \tilde{b}^*(f) + \tilde{a}^*(f) \tilde{b}(f) \right],
\end{equation}
where $S_h(f)$ is the (real) one-sided noise spectral density of the antenna and $f_{\rm low}$ and 
$f_{\rm high}$ are some appropriately defined lower and upper frequency cutoffs depending on the
signal and/or the antenna response.

Starting from the covariance matrix one defines the matrix of correlation
coefficients $C_{kl}$ in the following manner: 
\begin{eqnarray}
C_{kl} & = & \sqrt {\Sigma_{kl}},\ \ {\rm if}\ \ k=l,\nonumber \\
       & = & \frac {\Sigma_{kl}}{\sqrt{\Sigma_{kk}\Sigma_{ll}}},\ \ {\rm if}\ \ k\ne l.
\end{eqnarray}
As is obvious the diagonal element $C_{kk}$ of the matrix of correlation coefficients
is the (one-sigma) standard deviation in the measurement of the parameter $\lambda_k,$ and
the $kl$-off-diagonal element $C_{kl}$ is the normalized covariance between the parameters
$\lambda_k$ and $\lambda_l.$ The correlation coefficients $C_{kl}$ take values in the range $[-1,1].$
A correlation coefficient $C_{kl}$ close to 1 (or $-1$) indicates that the parameters 
$\lambda_k$ and $\lambda_l$ are perfectly correlated (or anti-correlated) with each other.
If on the other hand the value is close to zero then there is little correlation 
between the two parameters.  When two parameters have a correlation coefficient 
close to $\pm 1$ then varying one of the parameters is equivalent to varying the 
other. In other words one doesn't produce two independent signal shapes by varying both
the parameters, varying only one of them should be sufficient. However, when two parameters
are not correlated then varying one of them produces a distinct shape as compared to
varying the other. Therefore, we can look at the matrix of correlation coefficients
to determine how many parameters are really independent.

Finally, we can diagonalize the covariance matrix to find the
{\it principal components.}  On diagonalization one has (linearly) transformed 
from the parameter set $\lambda_k$ to a new set $\mu_k$ effectively removing
in the process all the
covariances between the parameters. The principal components of the signal
are those transformed parameters that correspond to the largest diagonal elements.
Since we have (by construction) a symmetric matrix the eigenvalues are all real. 
Moreover, since the covariance matrix is the inverse of the metric the principal 
components are the inverses of the eigenvalues of the metric.  
Before we embark upon a discussion of the results let us look at two simple examples.

Our first example is a simple mathematical example: Consider a two-dimensional 
matrix $C_{kl}.$ Suppose $C_{11}=C_{22}=1$ and $C_{12}=C_{21}=\epsilon < 1.$
The eigenvalues $\lambda_\pm$ of such a matrix are $\lambda_\pm = 1\pm \epsilon.$
If $\epsilon$ is close to zero then both the eigenvalues are roughly the
same and both the parameters are equally important. If, however, $|\epsilon|$ is
close to 1 then one of the eigenvalues is nearly {\em zero} making the
corresponding transformed parameter unimportant. Thus, closer the correlation
coefficient is to $\pm 1,$ smaller is the significance of the parameter.

As a second example let us consider a sinusoidal signal 
$h(t; A, t_0, \varphi_0)= A \cos[\omega(t-t_0) + \varphi_0].$ Here
the parameters $A$ and $t_0$ are uncorrelated and so are $A$ and $\varphi_0,$ while $t_0$
and $\varphi_0$ are perfectly anti-correlated. Increasing $\varphi_0$ is 
equivalent to decreasing $t_0$ but the signal has really only two independent 
parameters leading to two principal components.

\section{Correlation coefficients for smirches}
We now apply the formalism developed in the previous section to the gravitational
wave signals produced by stellar mass black holes falling into supermassive
black holes. We first compute the covariance matrix and then diagonalize it
to obtain the principal components. 

Let us first examine the matrix of correlation coefficients. In Tables
1-3 we have displayed the correlation coefficients of the smirches 
for three different values of the mass ratio, $\eta=10^{-3},\, 10^{-4},\, 10^{-5}.$ 
We have also changed the orientations of the spin, orbital angular momentum
and source direction from one example to the other to get an idea of how
the correlation coefficients depend on the angular parameters. Tables 1-3
correspond to cases A, B and C, respectively;
in all cases the total mass $m=10^6 M_\odot,$ $a_1=0.9$ and the initial
orientation of the orbit is the same $(\theta_L,\varphi_L) = (\pi/2, \pi/3).$
\begin{center}
\begin{tabular}{cccccc}
              & $\eta$    & $\theta_1$ & $\varphi_1$ & $\theta_S$ & $\varphi_S$\\
{\bf Case A:} & $10^{-3}$ & $\pi/4$    & $\pi/2$     &  $\pi/8$    & $\pi$ \\
{\bf Case B:} & $10^{-4}$ & $\pi/3$    & $\pi/4$     &  $\pi/2$    & $\pi/3$ \\ 
{\bf Case C:} & $10^{-5}$ & $\pi/4$    & $\pi/2$     &  $\pi/2$    & $\pi/3$ \\ 
\end{tabular}
\end{center}

The results reported here should be taken as a hint on what the general results are
going to be but we cannot be absolutely certain about any of the results.
We plan to report a fuller treatment of the problem with exhaustive Monte-Carlo 
simulations to validate our results.
\begin{table}
\caption
{Correlation coefficients and errors in the estimation of parameters for a 
smirch corresponding to the following parameters:
$m=10^{6}M_\odot, \, \eta=10^{-3}, \, a_1=0.9, \, \theta_1=\pi/4,\,  \varphi_1=\pi/2, \, \theta_L=\pi/5, 
\varphi_L=\pi/3,\, \theta_S=\pi/8, \, \varphi_S=\pi$ and $n_{\rm cyc}= 910$
}
\begin{center}
\begin{tabular}{lrrrrrrrrr}
\hline
$k$         & $m$ & $\eta$ & $a_1$ & $\theta_1$ & $\varphi_1$ & $\theta_L$ & $\varphi_L$ & $\theta_S$ & $\varphi_S$ \\
\hline
$m$         & $  0.0033 $ \\ 
$\eta$      & $ \mathbf{-0.9592} $ & $  0.0023 $ \\ 
$a_1$       & $ \mathbf{-0.9336} $ & $ \mathbf{ 0.8775} $ & $  0.0035 $ \\ 
$\theta_1$  & $  0.1516 $ & $ -0.1776 $ & $ -0.1301 $ & $  0.3270 $ \\ 
$\varphi_1$ & $ -0.0275 $ & $  0.0326 $ & $  0.0262 $ & $ \mathbf{ 0.7073} $ & $  1.8633 $ \\ 
$\theta_L$  & $  0.0771 $ & $ -0.0840 $ & $ -0.0973 $ & $ \mathbf{ 0.7127} $ & $  0.3228 $ & $  0.5821 $ \\ 
$\varphi_L$ & $ -0.0449 $ & $  0.0526 $ & $  0.0472 $ & $ \mathbf{ 0.6975} $ & $ \mathbf{ 0.9984} $ & $  0.2855 $ & $  2.036 $ \\ 
$\theta_S$  & $ -0.0002 $ & $ -0.00341 $ & $ -0.0101 $ & $ -0.2898 $ & $  0.0269 $ & $ \mathbf{-0.8667} $ & $  0.0690 $ & $  0.8522 $ \\ 
$\varphi_S$ & $ -0.0150 $ & $  0.0194 $ & $  0.0197 $ & $ \mathbf{ 0.8135} $ & $ \mathbf{ 0.9055} $ & $ \mathbf{ 0.6809} $ & $ \mathbf{ 0.8875} $ & $ -0.3925 $ & $  1.3879 $ \\ 
\hline
\end{tabular}
\end{center}
\end{table}

The diagonal elements in these Tables are the expected standard deviations in
the measurement of the corresponding parameters when the (amplitude) 
signal-to-noise ratio is 10. In the case of $m,$ $\eta$
and $s_1$ we have quoted the relative error (namely, $\sigma_k/\lambda_k$ where
$\sigma_k$ is the expected standard deviation and $\lambda_k$ is the true value
of the parameter) while the error in angles is the absolute error
expected in their measurements. Since the signal-to-noise ratio is
proportional to $\sqrt{\eta n_{\rm cyc}}$ we have
chosen the quantity $\sqrt{\eta n_{\rm cyc}}$ to be roughly the same in
the three examples\footnote{Indeed, as a rule of thumb all smirches with
the same total mass $m$ but different $\eta (\ll 1),$ that begin at 
the same frequency and coalesce within the
LISA band will have the same signal-to-noise ratio. This is because, if
the starting frequency is the same, then as $\eta$ is lowered by a certain
factor the number of cycles, and the duration of the signal,
goes up by the same factor so that $\eta n_{\rm cyc}$ is the same. Even
though the duration of the signal changes quite a lot the frequency content
of the signal doesn't change so much during this interval as a result of which
the signal-to-noise ratio is roughly invariant. For ground-based sources such
a rule of thumb doesn't apply since these sources span the entire detector
band and the signal-to-noise ratio depends not only on the factor $\eta n_{\rm cyc}$
but also on where in the frequency band those cycles are located.}. 
For this reason, though our sources are all at the same distance
they produce similar signal-to-noise ratios. In reality they have to be
integrated over different time scales, the duration being proportional to $\eta^{-1}.$
Finally, note that though we have quoted
a total mass (essentially the mass of the supermassive black hole) of 
$10^6 M_\odot$ the results of this analysis hold good for other masses
as well provided the mass ratios are the same. The spectral density of
noise really doesn't vary too much over the duration of a signal so that
the results discussed here are also valid for ground-based detectors with
some minor modifications.

\begin{table}
\caption
{Correlation coefficients and errors in the estimation of parameters for a 
smirch corresponding to the following parameters:
$m=10^6M_\odot, \, \eta=10^{-4}, \, a_1=0.9, \, \theta_1=\pi/3,\,  \varphi_1=3\pi/4, \, \theta_L=\pi/5, 
\varphi_L=\pi/3,\, \theta_S=\pi/2, \, \varphi_S=\pi/3$ and $n_{\rm cyc}= 9.1 \times 10^3$
}
\begin{center}
\begin{tabular}{lrrrrrrrrr}
\hline
$k$         & $m$ & $\eta$ & $a_1$ & $\theta_1$ & $\varphi_1$ & $\theta_L$ & $\varphi_L$ & $\theta_S$ & $\varphi_S$ \\
\hline
$m$         & $  0.0003 $ \\ 
$\eta$      & $ \mathbf{-0.9449} $ & $  0.0002 $ \\ 
$a_1$       & $ \mathbf{-0.9782} $ & $ \mathbf{ 0.9358} $ & $  0.0003 $ \\ 
$\theta_1$  & $  0.0057 $ & $ -0.0068 $ & $ -0.0184 $ & $  0.5922 $ \\ 
$\varphi_1$ & $  0.0094 $ & $ -0.0113 $ & $ -0.0091 $ & $ \mathbf{ 0.6862} $ & $  0.8631 $ \\ 
$\theta_L$  & $  0.0087 $ & $ -0.0160 $ & $ -0.0169 $ & $ \mathbf{-0.9847} $ & $ \mathbf{-0.7009} $ & $  1.0201 $ \\ 
$\varphi_L$ & $  0.0200 $ & $ -0.0263 $ & $ -0.0331 $ & $ \mathbf{ 0.8773} $ & $ \mathbf{ 0.9425} $ & $ \mathbf{-0.8675} $ & $  1.2781 $ \\ 
$\theta_S$  & $  0.0096 $ & $ -0.0140 $ & $  0.0014 $ & $ \mathbf{-0.9954} $ & $ \mathbf{-0.6963} $ & $ \mathbf{ 0.9935} $ & $ \mathbf{-0.8761} $ & $  1.0203 $ \\ 
$\varphi_S$ & $ -0.0082 $ & $  0.0100 $ & $  0.0106 $ & $ -0.1538 $ & $ \mathbf{ 0.5862} $ & $  0.1540 $ & $  0.3332 $ & $  0.1550 $ & $  0.6145 $ \\ 
\hline
\end{tabular}
\end{center}
\end{table}

The values of the correlation coefficients whose absolute value is greater than 
0.5 is shown in bold-face to easily identify which parameters are correlated
and which aren't. A correlation coefficient of 0.5 does not mean that one of
the parameters can be completely ignored. However, it does mean that the two
parameters are not completely independent and we can save some computational
cost by varying a suitable combination of the two. If the correlation is close
to 1, however, we could save a lot in computational costs. 

In respect of correlation coefficients the parameters fall into two categories:
the angular parameters describing the orientation of the black hole's spin,
orbital angular momentum and source location fall into one category and 
the non-angular parameters corresponding to the magnitude of the massive black 
hole's spin and the two masses fall into another category. 
The parameters within each set are correlated with one another but
there is little correlation between parameters from the two categories.

Let us return to the examples given in Tables 1-3. 
The parameters $m,$ $\eta$ and $s_1$ are so strongly correlated that only 
one of these three parameters suffices in a search. Similarly, looking at the 
correlation between the angular parameters we see that just two or three of the
angular parameters might be good enough in a search. In the case of the
masses and spin magnitude the correlation coefficients remain roughly the same
in the three different examples.  However, this is not so in the case of 
angular parameters. For the sake of definiteness we declare a parameter to be 
independent, and would therefore be needed in a search, if its correlation coefficient
with the other parameters is smaller than  0.85.  Comparing Tables 1-3 we find that in the
first example (Table 1) $\theta_1,$ $\varphi_1,$ and $\theta_L$ are the three
independent angular parameters while in the second example (Table 2) $\theta_1,$
$\varphi_1,$ and $\varphi_S$ are the three independent parameters and in the third
example $\theta_1$ and $\varphi_S$ are the independent parameters. If the
independent parameters had remained the same then the search algorithm would 
be a lot simpler; we could determine forever the independent parameters and carry
out a search in that space irrespective of where in the parameter space we are.
However, now the independent parameters for our search (in a small region of the
parameter space) depends on where in the parameter space we are. One should measure
the matrix of correlation coefficients at a given point in the parameter space before 
embarking upon laying a grid of templates.

\begin{table}
\caption
{Correlation coefficients and errors in the estimation of parameters for a 
smirch corresponding to the following parameters:
$m=10^6M_\odot, \, \eta=10^{-5}, \, a_1=0.9, \, \theta_1=\pi/4,\,  \varphi_1=\pi/2, \, \theta_L=\pi/5, 
\varphi_L=\pi/3,\, \theta_S=\pi/2, \, \varphi_S=\pi/3$ and $n_{\rm cyc}= 9.1 \times 10^4$
}
\begin{center}
\begin{tabular}{lrrrrrrrrr}
\hline
$k$         & $m$ & $\eta$ & $a_1$ & $\theta_1$ & $\varphi_1$ & $\theta_L$ & $\varphi_L$ & $\theta_S$ & $\varphi_S$ \\
\hline
$m$         & $  2.5e-5$ \\ 
$\eta$      & $ \mathbf{-0.9370} $ & $  1.4e-5 $ \\ 
$a_1$       & $ \mathbf{-0.8804} $ & $ \mathbf{ 0.7873} $ & $  2.7e-5$ \\ 
$\theta_1$  & $ -0.0315 $ & $  0.0435 $ & $  0.0370 $ & $  0.4220 $ \\ 
$\varphi_1$ & $ -0.0003 $ & $  0.0018 $ & $ -0.0182 $ & $ \mathbf{-0.8500} $ & $  1.1942 $ \\ 
$\theta_L$  & $ -0.0124 $ & $  0.0188 $ & $ -0.0057 $ & $ \mathbf{ 0.9781} $ & $ \mathbf{-0.8812} $ & $  0.9363 $ \\ 
$\varphi_L$ & $ -0.0150 $ & $  0.0219 $ & $ -0.0047 $ & $ \mathbf{-0.7517} $ & $ \mathbf{ 0.9723} $ & $ \mathbf{-0.8242} $ & $  1.1773 $ \\ 
$\theta_S$  & $  0.0330 $ & $ -0.0490 $ & $ -0.0097 $ & $ \mathbf{ 0.9186} $ & $ \mathbf{-0.8790} $ & $ \mathbf{ 0.9733} $ & $ \mathbf{-0.8637} $ & $  0.9398 $ \\ 
$\varphi_S$ & $  0.0040 $ & $ -0.0063 $ & $ -0.0010 $ & $  0.1735 $ & $  0.2853 $ & $  0.1819 $ & $  0.2969 $ & $  0.1847 $ & $  0.5546 $ \\ 
\hline
\end{tabular}
\end{center}
\end{table}

\section{Principal components for smirches}
Next we turn to the principal components of the spin modulated chirp. 
Recollect that the principal components are nothing but the eigenvalues of
the covariance matrix. 
In Table 4 we have shown the principal components for the cases discussed
in Tables 1-3. We have displayed only those components that are a significant
fraction of the largest component. (A rigorous treatment of the problem should
deal with dimensionless parameters so that in the process of diagonalization the
transformation of the parameters is meaningful. We defer such a treatment to
a later paper.)  We note that  the number of principal components is much 
less than the number of parameters although the precise number of principal
components depends on the point in the parameter space.  Our preliminary 
conclusion is that the problem of catching smirches boils down to searching 
for three, or at most four, parameters.  This means that we might need a 
lot fewer templates than was originally conceived. After all the problem 
may really not be as serious as we had originally thought.
\begin{table}
\caption
{Principal components for the three cases discussed in Tables 1-3. Only components
greater than $10^{-2}$ (chosen in an ad hoc manner) are listed. Cases A-C correspond
to the systems discussed in Tables 1-3 (see text).} 
\begin{center}
\begin{tabular}{ccc}
\hline
\hline
Case A          & Case B        & Case C\\
\hline
 13            & 445	        & 398 \\
 9.3           & 4.3           & 4.3 \\
 0.19          & 0.025         & 0.11 \\
 0.15          & ---           & 0.016 \\
 0.12          & ---           & --- \\
 0.050         & ---           & --- \\
\hline
\hline
\end{tabular}
\end{center}
\end{table}

\section{Conclusions}
Gravitational waves emitted by stellar mass black holes falling into a supermassive
black hole encode a lot of information about the structure of the spacetime geometry
around the massive Kerr black hole. Decoding this information requires not only
accurate modelling of the emitted waveform but algorithms that can efficiently
dig out and discriminate these signals. Since the parameter space corresponding to
such systems is quite large it has been thought that matched filtering, which is
not only effective in detecting signals but also deciphering their information content,
would be too expensive in a search for these signals. 
In this paper we have explored the signal manifold corresponding to spin modulated
chirps, or {\it smirches} as we would like to call them, using the covariance matrix 
and the method of principal components. We conclude that the problem might be tractable
after all since the effective dimensionality of the parameter space might be a lot
smaller than what has been thought before.

The signal emitted by a small black hole in quasi-circular orbit around a massive
Kerr black hole is, in general, characterized by nine search parameters 
-- the two masses, the magnitude of the black hole's spin and
its orientation, initial orientation of the orbit and the direction to the source.
Our analysis has shown strong correlations between the two masses and the spin
magnitude as well as between the different angular parameters. 
We have found strong indications that matched filtering a short duration
smirch (1,000 to 100,000 cycles) might require a search in only only 
3-, or at most 4-dimensional parameter space. This means that one might be able
to use matched filtering to fit the strongest smirches in LISA's data and dig deeper 
for weaker signals. Before we can be completely optimistic about the result, a Monte-Carlo
simulation should be performed to see if the overlap of arbitrary smirches can
be larger than a certain minimal match (say 50\%) when maximized over just three
parameters.

\ack
We are grateful to Sam Finn and Penn Statue University for hospitality
during the LISA symposium.  For support received BSS wishes to 
thank the Max Planck Institute for Gravitational Physics (Albert Einstein 
Institute), Golm, where this work was 
initiated.  This research was partly supported by the Particle 
Physics and Astronomy Research Council of the United Kingdom.

\end{document}